# Analysis of Frequency-Agile CSMA Wireless Networks

Soung Chang Liew*, Jialiang Zhang*, Chi-Kin Chau[+], Minghua Chen*

*Department of Information Engineering
The Chinese University of Hong Kong
New Territories, HK SAR, China
{soung, jlzhang4, minghua}@ie.cuhk.edu.hk

[+] Computer Laboratory
University of Cambridge
United Kingdom
Chi-Kin.Chau@cl.cam.ac.uk

*Abstract*—This paper proposes and analyzes the performance of a simple frequency-agile CSMA MAC protocol. In this MAC, a node carrier-senses multiple frequency channels simultaneously, and it takes the first opportunity to transmit on any one of the channels when allowed by the CSMA backoff mechanism. We show that the frequency-agile MAC can effectively 1) boost throughput and 2) remove temporal starvation. Furthermore, the MAC can be implemented on the existing multiple-frequency setup in Wi-Fi using multi-radio technology, and it can co-exist with the legacy MAC using single radio. This paper provides exact stationary throughput analysis for regular 1D and thin-strip 2D CSMA networks using an "transfer-matrix" approach. In addition, accurate approximations are given for 2D grid networks. Our closed-form formulas accurately quantify the throughput gain of frequency-agile CSMA. To characterize temporal starvation, we use the metric of "mean residual access time" (MRAT). Our simulations and closed-form approximations indicate that the frequency-agile MAC can totally eliminate temporal starvation in 2D grid networks, reducing its MRAT by orders of magnitude. Finally, this paper presents a "coloring theorem" to justify the use of the frequency-agile MAC in general network topologies. Our analysis and theorem suggest that with enough frequency channels, the frequency-agile MAC can effectively decouple the detrimental interactions between neighboring links responsible for low throughput and starvation.

*Index Terms* – CSMA; Wi-Fi; IEEE 802.11; Frequency Diversity.

## I. INTRODUCTION

It is common today to find multiple wireless LANs co-located in the neighborhood of each other. These wireless LANs form an overall large network. Their links interact and compete for airtime using the carrier-sense multiple access (CSMA) protocol. The carrier sensing relationships among these links are subtle in that each link may sense only a subset, but not all, of other links. It is known that links in such a network may suffer from 1) low equilibrium throughput, and 2) temporal starvation [1].

This paper is an investigation on how multiple frequency channels can be used to solve these two problems. Although the use of multiple channels in CSMA has been explored before [2-11], this paper makes new contributions on both design and analysis fronts. On the design front, we propose a simple frequency-agile MAC protocol to allow links in the network to exploit frequency diversity in an adaptive and distributed manner. This MAC protocol extends the traditional CSMA MAC protocol so that a node monitors and carrier-senses multiple frequency channels at the same time, and takes the first opportunity to transmit on any one of the channels when allowed by the CSMA backoff mechanism.

On the analysis front, we present closed-form results and a general coloring theorem showing that the frequency-agile MAC can boost throughput and remove temporal starvation effectively. To obtain concrete closed-form results, this paper focuses on regular networks. We study networks that can be modeled by 1D linear contention graphs and 2D grid contention graphs. By means of a "transfer-matrix" analytical approach from statistical physics, we obtain closed-form equations for the equilibrium throughputs of the 1D case with and without frequency diversity. Prior work [12] has studied the 1D case without frequency diversity; however, there were no closed-form solutions. The transfer-matrix approach can also be used to analyze "thin-strip" 2D regular networks in which one dimension is much larger than the other dimension.

The 2D case in which both dimensions are large and of similar size is a tough problem. We show that the CSMA network is equivalent to the Ising model with external magnetic field in statistical physics, a well-known hard problem that has defied closed-form solutions for decades. However, we manage to obtain highly accurate approximate solutions in the context of CSMA networks.

Links in CSMA networks without frequency diversity are prone to equilibrium and temporal starvations. "Temporal" starvation is to be distinguished from "equilibrium" starvation. A link suffering from equilibrium starvation has near-zero equilibrium throughput. In that respect, the part of this paper that addresses low equilibrium throughput also addresses equilibrium starvation. Most prior works on starvation in CSMA networks focus on equilibrium starvation. Ref. [6], for example, explored the use of frequency diversity to mitigate equilibrium starvation. Temporal starvation is a phenomenon in which a link may have acceptable equilibrium throughput, but yet may receive near-zero throughput for very long durations. That is, the link's throughput is highly uneven over time. Refs. [1] and [12] touched on concepts such as "island states" and "phase transition" to explain the underlying cause for temporal starvation in the 2D grid without frequency diversity.

In this paper, we propose to use the metric of mean residual access time (MRAT) to characterize temporal starvation. Without frequency diversity, the MRAT of a link in the 2D grid can be up to several thousands of packet transmission times. With frequency diversity, our frequency-agile MAC can reduce MRAT by orders of magnitude. In particular, each link behaves as if it were an isolated link free from the detrimental coupling effects from neighbor links. We believe the observation that frequency diversity, when exploited properly, can remove temporal starvation is new, and it provides another motivation for frequency-diverse CSMA networks besides boosting throughput.

Last but not least, to demonstrate the capability of our frequency-agile MAC in networks with general topologies, we prove a coloring theorem. We show that given a network with a contention graph that is $q$-colorable, if there are $q$ frequency channels, then our frequency-agile MAC allows each link to achieve a normalized throughput of one and there will be no temporal starvation. Notable is the fact that the frequency-agile MAC does not make use of an explicit coloring algorithm to achieve this desirable property. It is distributed and can adapt quickly to topological changes without explicit knowledge of the network topology. Although the proof of the theorem is simple, it relies on a non-trivial result that underlying the frequency-agile MAC is a time-reversible dynamic, to which there is a simple product-form solution for the stationary probability.

The remainder of the paper is organized as follows. Section II introduces our system model, presents our frequency-agile MAC, and reviews the basic equations governing CSMA networks. Sections III and IV analyze 1D and 2D regular networks. Section V presents the performance results obtained from analysis as well as simulations. Section VI proves the coloring theorem. Section VII provides an overview of related work. Section VIII concludes this paper.

## II. SYSTEM MODEL AND BASIC EQUATIONS

In Part A, we first review an idealized version of the CSMA network (ICN) that captures the main features of the CSMA protocol responsible for the interaction and dependency among links. The ICN model was used in several prior investigations [1] [12-14]. The correspondence between ICN and the IEEE 802.11 protocol [15] can be found in [1]. An important aspect of ICN is that the stationary probability of the system state is in a product form that provides much analytical and design insight. In Part B, we introduce our frequency-agile MAC. In Part C, we extend the product-form ICN equation for the stationary probability of frequency-agile CSMA networks.

To limit our scope, this paper focuses on one-hop networks. We choose this focus because the one-hop case is the most common way CSMA wireless networks (i.e., Wi-Fi) are deployed in practice, and the one-hop analysis can be extended for the multi-hop case.

### A. The ICN model

In ICN, the carrier-sensing relationships among links are described by a contention graph $G = (V, E)$. Each link is modeled as a vertex $i \in V$. This paper uses the terms "links" and "vertices" interchangeably. There is an edge $e \in E$ between two vertices if the transmitters of the two links can sense each other.

At any time, a link is either active or idle. A link is active when data is transmitted between its two end nodes. Thanks to carrier sensing, any two links that can sense each other will refrain from being active at the same time. A link sees the channel as idle if and only if none of its neighbors is active.

In the ICN model, each link maintains a *backoff timer* $C$, the initial value of which is a random variable with an *arbitrary* distribution $f(t_{cd})$ with mean $E[t_{tr}]$. The timer decreases in value in a continuous manner with $dC/dt = -1$ as long as the link senses the channel as idle. If the channel is sensed busy, the countdown process is frozen and $dC/dt = 0$. When the channel becomes idle again, the countdown continues and $dC/dt = -1$ with $C$ initialized to the previous frozen value. When $C$ reaches zero, the link transmits a packet. The packet transmission time is a random variable with *arbitrary* distribution $g(t_{tr})$ with mean $E[t_{cd}]$. After the transmission, the link resets $C$ to a new random value according to the distribution $f(t_{cd})$, and the process repeats.

Let $s_i \in \{0,1\}$ denote the state of link $i$, where $s_i = 1$ if link $i$ is active (transmitting) and $s_i = 0$ if link $i$ is idle (actively counting down or frozen). The overall system state is $s = (s_1, s_2, ..., s_N)$, where $N$ is the number of links in the network. Note that $s_i$ and $s_j$ cannot both be 1 at the same time if links $i$ and $j$ are neighbors because they can sense each other. The collection of feasible states corresponds to the collection of independent sets (IS) [16] of the contention graph.

As an example, Fig. 1(a) shows a contention graph with four links, and Fig. 1(b) shows the corresponding state-transition diagram. As a mental picture to map the scenario to Wi-Fi, the reader could imagine each vertex as consisting of an AP and a client station, and that there are four WLANs in this example. In Fig. 1(b), to avoid clutters, we merge the two directional transitions between two states into one line. Each transition from left to right corresponds to the beginning of the transmission of one particular link, while the reverse transition corresponds to the ending of the transmission of that link.

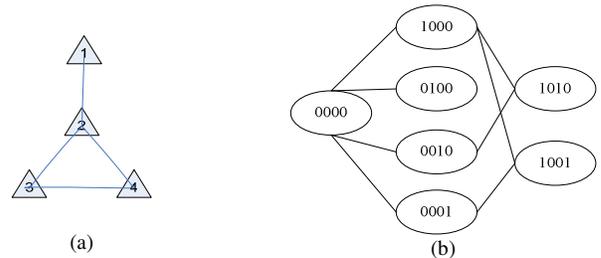

Fig. 1. (a) An example contention graph and (b) its state-transition diagram.

*Equilibrium analysis of ICN*

If we further assume that $f(t_{cd})$ and $g(t_{tr})$ are exponential distributions, then it can be shown that $s(t)$ is a time-reversible Markov process [1]. Thus, detailed balance applies to any two neighbor states (e.g., $P_{1010}/E[t_{tr}] = P_{1000}/E[t_{cd}]$ in Fig. 1(b)). Applying detailed balance to the overall Markov chain gives the following stationary probability distribution [1]:

$$P_s = \frac{\rho^{n_s}}{Z} \; \forall \, s \in \mathcal{S}, \quad \text{where } Z = \sum_{s \in \mathcal{S}} \rho^{n_s} \quad (1)$$

where $\rho = E[t_{tr}]/E[t_{cd}]$; $\mathcal{S}$ is the set of all feasible states;

and $n_s$ is the number of ongoing transmissions in the network when the system is in state $s = (s_1, s_2, ..., s_N) \in \mathcal{S}$. For example, in Fig. 1, $n_{1010} = 2$. In this paper, we refer to $\rho$ as the *access intensity* of links. We define the normalized throughput of link $i$ to be the fraction of time during which it is transmitting: $th_i = \sum_{s:s_i=1} P_s$.

Ref. [1] showed that (1) is in fact quite general and does not require $s(t)$ to be a strict Markov process. In particular, (1) is insensitive to the distributions $f(t_{cd})$ and $g(t_{tr})$ given the ratio of their mean $\rho$.

### B. Frequency-Agile ICN

In this paper, we consider frequency-agile CSMA networks in which multiple frequency channels are available. We introduce a simple frequency-agile MAC and then extend (1) for the stationary probability distribution in networks using the MAC.

*Operation of $(q,k)$ ICN*

In a $(q,k)$ network, there are $q$ available frequency channels, and each link can transmit on at most $k \leq q$ channels at a given time. Each link has $q$ backoff timers, $C_1, ..., C_q$, one for each channel. Timer $C_f$ controls the access to channel $f$. When no neighbor transmits on channel $f$, $C_f$ decreases in value with $dC_f/dt = -1$; when a neighbor transmits on channel $f$, $C_f$ freezes; in addition, if the link is already transmitting on $k$ channels, $C_f \forall f$ also freeze. When $C_f$ reaches zero, the link transmits a packet on channel $f$ with duration following distribution $g(t_{tr})$; when the transmission completes, the link resets $C_f$ to a new random value according to distribution $f(t_{cd})$, and the process repeats. □

In Wi-Fi, a number of orthogonal frequency channels are available. The frequency-agile MAC can operate over them using multi-radio technology. Also, the frequency-agile MAC is compatible and can co-exist with the traditional non-frequency agile MAC. In particular, MACs that access different numbers of frequency channels can co-exist.

The following describes how (1) can be generalized to the $(q,1)$ case. Generalization to the $(q,k)$ case is similar after proper redefinition of the link state $s_i$. In the $(q,1)$ case, a link can transmit on at most one channel at a time. The state of a link $i$ is $s_i \in \{0, 1, ..., q\}$, where $s_i = 0$ if link $i$ is not transmitting, and $s_i = f$ if link $i$ is transmitting on channel $f$. The overall system state is $s = (s_1, s_2, ..., s_N)$.

As in the (1,1) case, the process is time-reversible and the argument as in [1] can be applied to obtain the same product-form solution as in (1), but with expanded feasible states $\mathcal{S}$. For example, for the contention graph in Fig. 2(a), when $(q,k) = (1,1)$, $\mathcal{S} = \{000, 001, 010, 100\}$. When $(q,k) = (2,1)$, $\mathcal{S}$ contains many more states, and the state-transition diagram is shown in Fig. 2(b).

Although the analytical techniques in this paper can be applied to the general $(q,k)$ case, we will focus on the $(q,1)$ case. There are two reasons for our focus: 1) As will be shown, the $(q,1)$ case is quite effective in solving throughput and starvation problems; 2) the physical layer is simpler to implement when $k = 1$, as explained below. In CSMA networks such as Wi-Fi, each DATA transmission is followed by an ACK. Suppose that $k = 2$. Then, it is possible for a node to be transmitting DATA on two channels at the same time. If one DATA transmission finishes before the other, the node will be transmitting DATA on one channel while receiving ACK on the other channel. The very strong signal of the DATA transmission could cause the ACK reception to be corrupted due to out-of-band leakage from the transmission band to the reception band (see [17] for more details).

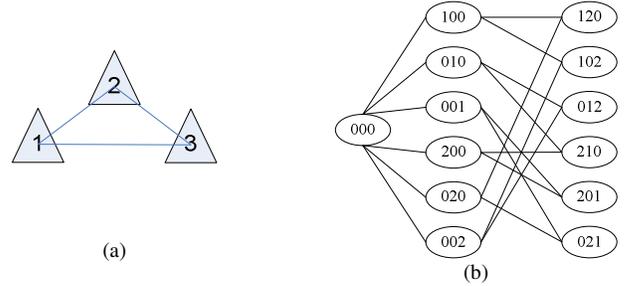

Fig. 2. (a) An example contention graph and (b) its state-transition diagram when $(q,k) = (2,1)$.

### C. Compatibility Function, Partitoin Fuction, and Relationship with Throughput in Regular CSMA Networks

For the transfer-matrix analysis, it is more convenient to write (1) in a different form, in which the dependency between the states of two neighbor vertices, $s_i$ and $s_j$, are captured by a compatibility function, $\psi_{ij}(s_i, s_j) \in \{0, 1\}$. Specifically, $\forall (i, j) \in E$, $\psi_{ij}(s_i, s_j) = 1$ if $s_i$ and $s_j$ are compatible, and $\psi_{ij}(s_i, s_j) = 0$ if they are not. For example, for $(q,k) = (2,1)$, two neighbors cannot transmit on the same frequency. Thus, $\forall (i, j) \in E$, $\psi_{ij}(s_i, s_j) = 0$ if $s_i = s_j = 1$ or $s_i = s_j = 2$; $\psi_{ij}(s_i, s_j) = 1$ otherwise.

For the general $(q,k)$ case, let $\mathcal{X}$ denote all possible states of a vertex. We can rewrite (1) as

$$P_s = \frac{\prod_{(i,j) \in E} \psi_{ij}(s_i, s_j) \prod_{i \in V} \phi_i(s_i)}{Z} \quad \forall s \in \mathcal{X}^N;$$
$$Z = \sum_{s \in \mathcal{X}^N} \prod_{(i,j) \in E} \psi_{ij}(s_i, s_j) \prod_{i \in V} \phi_i(s_i) \quad (2)$$

where $\phi_i(s_i) = \rho^{n_{s_i}}$, $n_{s_i}$ is the number of frequency channels on which link $i$ is transmitting when in state $s_i$; $n_s = \sum_{s_i} n_{s_i}$. For example, for a $(2,1)$ network, $\phi_i(0) = 1$; $\phi_i(1) = \phi_i(2) = \rho$. In general, for a $(q,1)$ network, $\phi_i(s_i) = 1$ if $s_i = 0$; $\phi_i(s_i) = \rho$ if $s_i \neq 0$.

Note that in (2), unlike in (1), $s \in \mathcal{X}^N$ and $s$ can be feasi-ble as well as infeasible. If it is infeasible, then $P_s = 0$ because a compatibility function in the numerator will be zero. Thus, (2) is consistent with (1).

The normalization factor $Z$ (also called the partition function) contains much information about the system performance. In particular, the total system is related to $Z$ as follows:

$$Th \triangleq E[n_s] = \rho \frac{d \ln Z}{d\rho} \quad (3)$$

To see the above, note from (1) that $d \ln Z / d\rho = \left(\sum_{s \in \mathcal{S}} n_s \rho^{n_s - 1}\right) / Z = \left(\sum_{s \in \mathcal{S}} n_s \rho^{n_s} / Z\right) / \rho = E[n_s]/\rho$.

## III. 1-D NETWORKS

This section is devoted to the analysis of 1-D networks by a "transfer-matrix" approach. The gist of the transfer-matrix analysis is to derive the partition function $Z$ and then use (3) to obtain the throughput. For regular 1D networks (and thin-strip 2D networks in the next section), $Z$ is of the form $Z = \sum_i z_i^N$, where $z_i$ are the eigenvalues of a transfer matrix. Typically, $Z$ is dominated by the largest eigenvalue when the number of links $N$ is not too small. Thus, the whole problem becomes identifying the transfer matrix that characterizes the system and then finding its dominant eigenvalue.

The main body of this paper focuses on ring networks. A slight modification of the analysis allows it to be applied to open-ended linear networks (see Appendix A). In general, the vertices in the ring network behave like the internal vertices (those not near the boundaries) of the open-ended linear network. Thus, understanding the ring network allows us to understand the open-ended linear network as well.

We remind the reader that this paper focuses on one-hop networks. Each vertex models a one-hop link. As a mental picture, the reader could associate each vertex with a pair of nodes consisting of an AP and a client station.

### A. 1-D Ring Network: (1,1) Case

In a ring contention graph with $N$ vertices, each vertex has two neighbors. The neighbors of vertex $i$ are vertices $i-1$ and $i+1$. For convenience, notionally, vertices $N+1$ and $1$ refer to the same vertex. Then, (2) can be written as

$$P_{(s_1, s_2, \ldots, s_N)} = \frac{\prod_{i=1}^{N} \psi_{i,i+1}(s_i, s_{i+1}) \phi_i(s_i)}{Z}; \quad \forall (s_1, \ldots, s_N) \in \mathcal{X}^N$$

$$Z = \sum_{s_1, s_2, \ldots, s_N} \prod_{i=1}^{N} \psi_{i,i+1}(s_i, s_{i+1}) \phi_i(s_i) \quad (4)$$

For $(q,k) = (1,1)$, $\mathcal{X} = \{0,1\}$. In transfer-matrix analysis, we represent state $s_i$ by a vector, where $s_i = (1 \ 0)^T$ corresponds to $s_i = 0$, and $s_i = (0 \ 1)^T$ corresponds to $s_i = 1$. With this representation, we can write the term $\psi(s_i, s_{i+1})\phi_i(s_i)$ as

$$\psi(s_i, s_{i+1})\phi_i(s_i) = s_i^T \mathbf{A} s_{i+1};$$
$$\mathbf{A} = \begin{pmatrix} 1 & 1 \\ \rho & 0 \end{pmatrix} \quad (5)$$

$\mathbf{A}$ is a coupling matrix that expresses the dependency of $s_i$ on $s_{i+1}$. Rows 1 and 2 of $\mathbf{A}$ correspond to $s_i = 0$ and $s_i = 1$, respectively; and columns 1 and 2 of $\mathbf{A}$ correspond to $s_{i+1} = 0$ and $s_{i+1} = 1$, respectively. Consider the probability $P_{(s_1, s_2, \ldots, s_N)}$ in (4). For a particular state $s = (s_1, s_2, \ldots, s_N)$, the values of $s_i$ and $s_{i+1}$ determine which of the four entries in $\mathbf{A}$ contributes to the numerator $\prod_{i=1}^{N} \psi_{i,i+1}(s_i, s_{i+1})\phi_i(s_i)$ as one factor of the product. In particular, entry $(s_i, s_{i+1})$ of $\mathbf{A}$ gives the contribution of vertex $i$ to the product. The contributions of all vertices yield the numerator of $P_{(s_1, s_2, \ldots, s_N)}$.

Applying (5) on the partition function in (4), we have $Z = \sum_{s_1, \ldots, s_N} \prod_{i=1}^{N} s_i^T \mathbf{A} s_{i+1} = \sum_{s_1, \ldots, s_N} s_1 \mathbf{A} s_2 s_2^T \cdots s_N s_N^T \mathbf{A} s_{N+1}$. Applying associative law and noting that $\sum_{s_i} s_i s_i^T = \begin{pmatrix} 1 & 0 \\ 0 & 1 \end{pmatrix} = I_2$, we get

$$Z = \sum_{s_1} s_1^T \mathbf{A}^N s_{N+1} = \sum_{s_1} s_1^T \mathbf{A}^N s_1 = \text{Tr } \mathbf{A}^N = z_1^N + z_2^N \quad (6)$$

where $z_1$ and $z_2$ are the eigenvalues of $\mathbf{A}$. By solving for the roots of the characteristic polynomial of $\mathbf{A}$, they can be found to be

$$z_1, z_2 = \frac{1 \pm \sqrt{1 + 4\rho}}{2} \quad (7)$$

For the ring network, by symmetry, all vertices have the same throughput. Since the total system throughput is given by (3), the throughput of a vertex $i$ is given by $Th_i = (\rho d \ln Z / d\rho)/N$. This yields

$$Th_i(\rho, N) = \frac{-z_2 z_1^N + z_1 z_2^N}{(z_1 - z_2)(z_1^N + z_2^N)} \quad \forall i \in \{1, \ldots, N\} \quad (8)$$

For large $N$, the larger eigenvalue $z_1$ dominates in $Z = z_1^N + z_2^N$, and $Z \approx z_1^N$. We have

$$Th_i = \frac{\rho}{N} \frac{d \ln Z}{d\rho} \approx \frac{\rho}{z_1} \frac{dz_1}{d\rho} \quad (9)$$

The formula in (9) will be used not just in the (1,1) case, but also in other cases. As will be seen in Section V, $N$ does not need to be very large for (9) to be accurate. Applying (7) on (9), for the (1,1) ring network, we obtain

$$Th_i(\rho, \infty) \approx \frac{\sqrt{1+4\rho} - 1}{2\sqrt{1+4\rho}} = \frac{1}{2} - \frac{1}{2\sqrt{1+4\rho}} \quad \forall i \in \{1, \ldots, N\} \quad (10)$$

### B. Frequency-Agile 1-D Ring: (2,1) Case

We now consider the frequency-agile 1-D ring network. We focus on the (2,1) case. Generalization to the $(q,k)$ case follows the same principle. For the (2,1) case, each vertex $i$ has three possible states: $s_i = (1\ 0\ 0)^T$ if link $i$ is idle; $s_i = (0\ 1\ 0)^T$ if link $i$ is transmitting on channel 1; and $s_i = (0\ 0\ 1)^T$ if link $i$ is transmitting on channel 2.[1] The compatibility function is encapsulated in the following transfer matrix:

$$\mathbf{A} = \begin{pmatrix} 1 & 1 & 1 \\ \rho & 0 & \rho \\ \rho & \rho & 0 \end{pmatrix} \quad (11)$$

The eigenvalues of $\mathbf{A}$ are given by the roots of the characteristic polynomial $z^3 - z^2 - (\rho^2 + 2\rho)z - \rho^2$:

$$z_1, z_2 = \frac{(\rho+1) \pm \sqrt{(\rho+1)^2 + 4\rho}}{2}, \quad z_3 = -\rho \quad (12)$$

Similar to the argument in the (1,1) case in (6), the partition function is found to be

$$Z = z_1^N + z_2^N + z_3^N \quad (13)$$

The throughput of a vertex $i$ is again given by $Th_i = (\rho\, d \ln Z / d\rho)/N$, which is

$$Th_i(\rho, N) = \frac{\frac{(z_1-1)z_1^N}{(z_1-z_2)} + \frac{(1-z_2)z_2^N}{(z_1-z_2)} + z_3^N}{z_1^N + z_2^N + z_3^N} \quad (14)$$

For large $N$, $z_1$ dominates in $Z$. Applying $z_1$ in (12) on (9), we get

$$Th_i(\rho, \infty) \approx \frac{\sqrt{(\rho+1)^2 + 4\rho} + \rho - 1}{2\sqrt{(\rho+1)^2 + 4\rho}} \quad (15)$$

For large $\rho$,

$$Th_i(\rho, \infty) \approx \frac{\rho}{\rho+1} = 1 - \frac{1}{\rho+1} \quad (16)$$

Note that the expression in (16) is also the throughput of an isolated link with no neighbors when $(q,k) = (1,1)$. Thus, when $\rho$ is large, having two frequency channels in the (2,1) ring network effectively decouples the vertices so that each vertex behaves like an isolated link. A second observation by comparing (10) and (16) is that the (2,1) system can roughly double the throughput of the (1,1) system for large $\rho$.

C. *L-nearest Neighbor Ring Network: Cases of (1,1), (2,1), (3,1)*

The above analysis can be extended to a ring network in which there is an edge between vertices $i$ and $j$ if they are $L$-closest neighbors. That is, if $(j-i) \bmod N \le L$.

(1,1) *Case:* The first step, as before, is to define the local state $s_i$. However, for $L > 1$, instead of just vertex $i$, $s_i$ represents the state of a unit consisting of vertices $i, ..., i+(L-1)$. At most one of these vertices can be transmitting at a given time. In vector form, $s_i$ is an $(L+1) \times 1$ vector. If all vertices $i, ..., i+(L-1)$ are idle, then $s_i = (1\ 0\ \cdots\ 0)^T$. If vertex $i+k$ is transmitting, then the $(k+2)^{th}$ element of $s_i$ is 1, and the other elements are 0.

The second step is to set up the transfer matrix, which can be found to be

$$\mathbf{A} = \begin{pmatrix} 1 & 0 & 0 & \cdots & 0 & 1 \\ \rho & 0 & 0 & \cdots & 0 & 0 \\ 0 & 1 & 0 & \cdots & 0 & 0 \\ 0 & 0 & 1 & \cdots & 0 & 0 \\ \vdots & \vdots & \vdots & \ddots & \vdots & \vdots \\ 0 & 0 & 0 & \cdots & 1 & 0 \end{pmatrix} \quad (17)$$

The characteristic polynomial of $\mathbf{A}$ is $z^{L+1} - z^L - \rho$. There are $L+1$ eigenvalues: $z_1, ..., z_{L+1}$. The partition function is

$$Z = z_1^N + \cdots + z_{L+1}^N \quad (18)$$

To find the throughput of a vertex $i$, we again use the formula $Th_i = (\rho\, d \ln Z/d\rho)/N$. It can be shown (omitted here to conserve space) from $z^{L+1} - z^L - \rho = 0$ that there is a unique real positive eigenvalue whose magnitude is larger than those of all other eigenvalues. Without loss of generality, let $z_1$ denote this dominant eigenvalue.

For large $N$, $Z \approx z_1^N$. Note that $z_1^{L+1} - z_1^L - \rho = 0$ by definition. Differentiating it gives $[(L+1)z_1^L - Lz_1^{L-1}]dz_1/d\rho - 1 = 0$. Using these two equations on (9) yields

$$Th_i(\rho, \infty) = \frac{1}{L+1} - \frac{1}{L+1} \cdot \frac{1}{(L+1)z_1 - L} \quad (19)$$

For $L = 2$, $z_1 \approx \rho^{1/3} + 1/3 + 1/(9\rho^{1/3})$. Thus,

$$Th_i(\rho, \infty) \approx \frac{1}{3} - \frac{1}{9\rho^{1/3} + 1/\rho^{1/3} - 3} \quad (20)$$

Using a similar approach, we can derive the vertex throughputs of the cases of (2,1) and (3,1).

(2,1) *Case:* For the (2,1) system with $L = 2$, the characteristic polynomial is $(z^3 + \rho z - \rho^2)[z^4 - z^3 - \rho z^2 - (\rho^2 + \rho)z - \rho^2]$. The dominant eigenvalue $z_1$ is approximately $\rho^{2/3} + \rho^{1/3}/3 + 2/3$. Using (9), we can obtain the following vertex throughput:

$$Th_i(\rho, \infty) \approx \frac{2}{3} - \frac{\rho^{1/3}/9 + 4/9}{\rho^{2/3} + \rho^{1/3}/3 + 2/3} \quad (21)$$

---
[1] For general $(q,k)$, $s_i$ has $\sum_{j=1}^{k} C_j^q$ possible states. We will need to represent $s_i$ as a vector of similar dimension.

(3,1) *Case:* For the (3,1) system with $L=2$, the characteristic polynomial is of the $13^{th}$ order. The dominant eigenvalue $z_1$ is a root of one of its factor polynomial $z^4 - (1+\rho)z^3 - \rho z^2 - (2\rho^2 + \rho)z - \rho^2$. Using (9), the vertex throughput for large $N$ can be found to be approximately

$$Th_i(\rho, \infty) \approx \frac{\rho(z_1^3 + z_1^2 + z_1 + 4\rho z_1 + 2\rho)}{z_1(4z_1^3 - 3\rho z_1^2 - 3z_1^2 - 2\rho z_1 - \rho - 2\rho^2)} \qquad (22)$$

Unfortunately, the Taylor expansion of $z_1$ in terms of $\rho$ does not converge smoothly even after the seventh term because of the large coefficients. Thus, in our numerical study, we find $z_1$ numerically for each $\rho$ and substitute that into (22) to get the vertex throughput.

## IV. 2-D NETWORKS

In this section, we applies the transfer-matrix approach to thin-strip 2D networks. We show that the 2D torus network is difficult to analyze exactly by establishing the correspondence between the CSMA network and the Ising model with external magnetic field in statistical physics. We then provide accurate approximate solutions to the 2D torus network.

### A. Thin-Strip Ring Networks

Thin-strip ring networks have structures as shown in Fig. 3(a). The structure is long horizontally and consists of units that couple with each other in a repetitive manner. Each unit is a local contention graph consisting of multiple vertices. The number of states in the local contention graph, $M$, is the number of states in the local unit. We can define $s_i$ to be an $(M+1) \times 1$ vector accordingly to capture the state of the local unit. The coupling between adjacent units are then represented by an $(M+1) \times (M+1)$ transfer matrix $\mathbf{A}$.

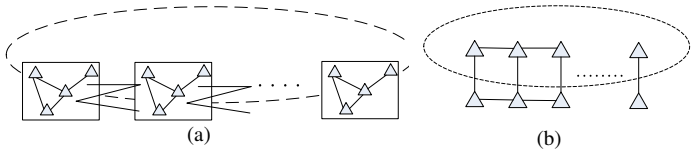

Fig. 3. (a) Structure of a thin-strip 2D network; (b) a particular thin-strip 2D network.

Fig. 3(b) shows an example of thin-strip networks. The local unit consists of just two connected vertices. For two adjacent units, there is an edge between the two top vertices, and an edge between the two bottom vertices. Appendix B presents the analysis of this network for $(q, k) = (1,1)$ and $(2,1)$.

### B. 2-D Torus Networks

All the 1D networks studied in this paper can be extended to 2D networks by adding another dimension. For example, to extend the 1D ring in Section III-A, we index a vertex by $(i, j)$ rather than just $i$. The state is $s_{ij}$. There will be altogether $N^2$ vertices rather than just $N$ vertices. Each vertex $(i, j)$ has four neighbors: $(i, j+1 \bmod N)$, $(i, j-1 \bmod N)$, $(i+1 \bmod N, j)$, and $(i-1 \bmod N, j)$.

It turns out the analysis for the 2D torus network is a tough problem. In Appendix C, we establish the correspondence between the 2D torus CSMA network and the 2D torus Ising model [18] with external magnetic field. The Ising model was invented to study ferromagnetism in statistical physics. The celebrated result by Onsager [18] in 1944 on 2D Ising model deals with the case in which there is no external magnetic field to the ferromagnetic material. To-date, there is no closed form solution for the case with external magnetic field. Unfortunately, the CSMA network case is one that can be mapped to the Ising model with external magnetic field.

For an analytical approximation for 2D torus, we consider the large $\rho$ case. Section V will present simulation results showing that our approximation is accurate even for modest-size $\rho$, such as that typical in standard Wi-Fi networks (for TCP/IP packets in 802.11b, $\rho \approx 5$).

(1,1) *Case:* When $\rho$ is large, the system has two possible phases [1][12]. In one phase, the odd vertices $(i, j)$, where $i+j = odd$, dominate; and in the other phase, the even vertices $(i, j)$, where $i+j = even$, dominate. When the odd vertices dominate, each even vertex behaves approximately like a link flanked by four isolated links, being the center of a star contention graph. Using (1), the even vertex throughput can be computed to be $\rho/(1 + 5\rho + 6\rho^2 + 4\rho^3 + \rho^4)$. This throughput is very small, and the odd vertices behave essentially like isolated links with throughput $\rho/(1+\rho)$. Over the long term, by symmetry, the odd and even vertices dominate with equal probability. Thus, the long-term vertex throughput can be approximated to be

$$Th_{ij}(\rho) \approx \frac{1}{2} \cdot \left( \frac{\rho}{1+\rho} + \frac{\rho}{1 + 5\rho + 6\rho^2 + 4\rho^3 + \rho^4} \right) \qquad (23)$$

(2,1) *ICN:* When there are two channels, the two phases take on a different form. In one phase, the odd vertices behave like isolated links transmitting on frequency $f_1$, and the even vertices behave like isolated links transmitting on frequency $f_2$. In the other phase, the odd and even vertices adopt opposite frequencies. The link throughput can be approximated by that of an isolated link operating on one frequency. Thus,

$$Th_{ij}(\rho) \approx \frac{\rho}{1+\rho} \qquad (24)$$

## V. PERFORMANCE RESULTS

This section examines the performance results from analysis as well as simulation. We focus on two performance measures: 1) throughput; 2) temporal starvation. We propose to use the "mean residual access time" (MRAT) to characterize the latter (to be elaborated in Part B).

The main results are as follows:
1. Our analytical throughput solutions that adopt large-$N$ approximation are very accurate even for modest-size $N$ (specifically, $N \geq 16$).
2. Link throughput increases roughly proportionally with

$q$ in $(q,1)$ regular networks when $q \leq C_G$, where $C_G$ is the chromatic number of the contention graph $G$.

3. MRAT in 1D regular networks are largely benign and there is no severe temporal starvation. However, MRAT in 2D torus without frequency diversity can be orders of magnitude higher than that of an isolated link. Fortunately, having $q = 2$ solves the problem completely.

We performed simulation experiments with MATLAB 7.0 on a Pentium(R) M 1.86GHz laptop. In our simulations, we assume $f(t_{cd})$ and $g(t_{tr})$ to be exponentially distributed, with means $E[t_{cd}] = 1/\rho$ and $E[t_{tr}] = 1$, respectively. That is, we normalize time so that one time unit is equal to one mean packet duration. The total simulation time is set to be more than one million time units. We computed the 90% confidence intervals to characterize the accuracy of our results.

*A. Throughput*

For ease of reference, Table 1 lists the equations that will be compared against our simulation results. Fig. 4 plots throughput versus $\rho$ for the ring networks with $L = 1$ and $(q,k) = (1,1), (2,1)$. The access intensities and network dimensions tested are $\rho = 5, 10, 15, 20$; $N = 16, 36, 64$. The approximations of (10) and (15) overlap with the simulated results of $(q,k) = (1,1)$ and $(2,1)$, respectively. The 90% confidence intervals are very small and hardly discernable. Note that the link throughputs in the $(2,1)$ systems roughly double the link throughputs of the corresponding $(1,1)$ systems.

Fig. 5 plots throughput versus $\rho$ for the ring networks with $L = 2$ and $(q,k) = (1,1), (2,1), (3,1)$. Again, our approximations in (20), (21), and (22) are rather accurate. Note that with $L = 2$, three colors are needed to color the contention graph. We see that the ratio of the throughputs of the (1,1), (2,1) and (3,1) systems are roughly 1:2:3.

Fig. 6 plots throughput versus $\rho$ for the 2D torus networks with $(q,k) = (1,1), (2,1)$. The network dimensions considered are $N = 4 \times 4, 6 \times 6, 8 \times 8$. Recall that exact analysis for this case is not amenable to closed-form solutions. Yet, our approximations in (23) and (24) are nearly perfect. The torus is a 2-colorable graph. We see that the throughput of the (2,1) case is roughly twice that of the (1,1) case.

To conserve space, we omit the plotting of the results of the thin-strip network of Fig. 3(b) here. The results are qualitatively the same as in the 1D networks, with the (2,1) system having nearly twice the throughput of the (1,1) system. Furthermore, our approximations are very tight.

Table 1. Analytical equations to be compared against simulation results.

| $(q,k)$ | 1D Ring, $L = 1$ | 1D Ring, $L = 2$ | 2D Torus |
|---|---|---|---|
| (1,1) | (10) | (20) | (23) |
| (2,1) | (15) | (21) | (24) |
| (3,1) | -- | (22) | -- |

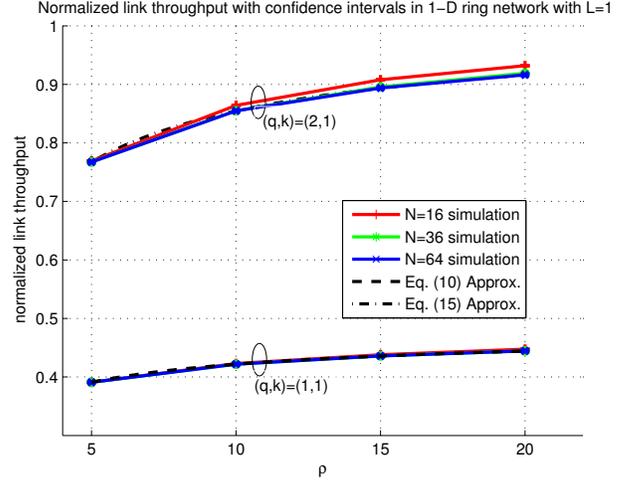

Fig. 4. Normalized link throughput versus access intensity for ring networks with $L = 1$ and $(q,k) = (1,1), (2,1)$.

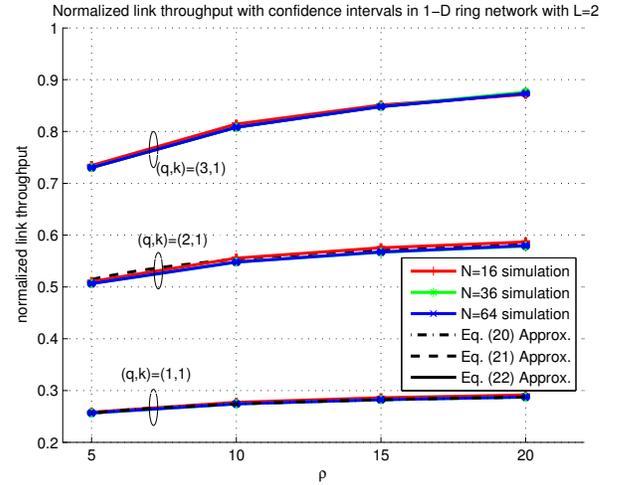

Fig. 5. Normalized link throughput versus access intensity for ring networks with $L = 2$ and $(q,k) = (1,1), (2,1), (3,1)$

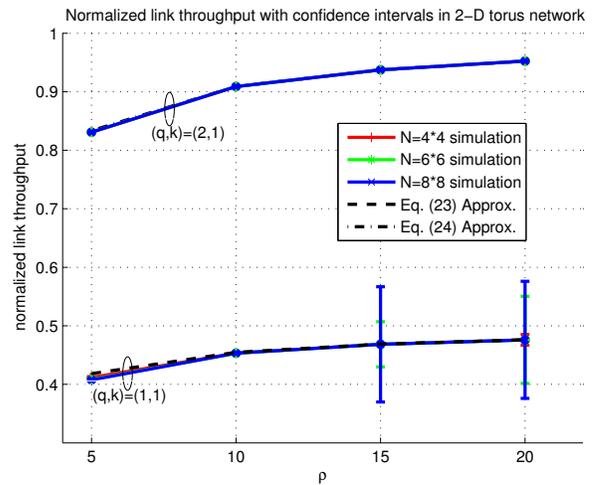

Fig. 6. Normalized link throughput versus access intensity for 2D torus networks with $(q,k) = (1,1), (2,1)$.

## B. Temporal Starvation

We next consider temporal starvation. We will see that our frequency-agile MAC can alleviate temporal starvation very effectively.

The throughputs studied in Part A are the equilibrium throughputs. Equilibrium throughput does not capture the temporal behavior of the system. Even if the equilibrium throughput is acceptable, a link may still undergo long durations during which it cannot transmit any packet at all.

Let us illustrate this with the 1D ring with 36 links, and 2D torus $6 \times 6$ links. Consider the case of $(q,k) = (1,1)$, $L = 1$, $\rho = 10$. According to Fig. 4 and Fig. 6, the equilibrium link throughputs for both networks are above 0.4, with the torus having slightly higher throughput than the ring. Fig. 7 (a) and (b) plot the temporal behavior of the two networks. For each network, the on-off pattern of one specific link is plotted.

The figure reveals that the ring does not suffer from temporal starvation, whereas the torus does. From Fig. 7(a), we see that a link in the ring may undergo an off period of up to around 10 time units (one unit = one mean packet transmission time). By contrast, from Fig. 7(b), we see that a link in the torus may undergo an off period of up to thousands of time units. During this time, the link is said to undergo temporal starvation.

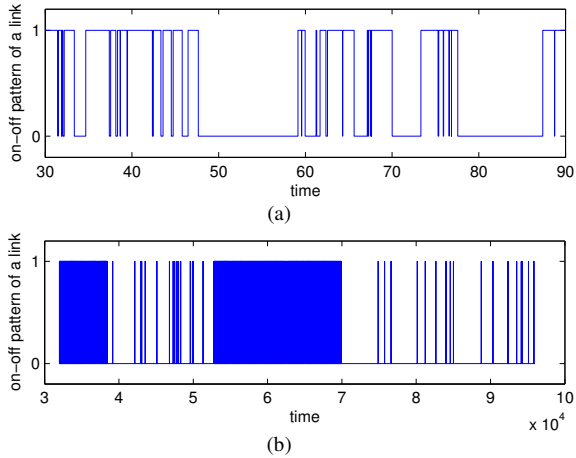

Fig. 7. On-Off transmission patterns of (a) a link in a ring network, and (b) a link in a torus network.

A measure to capture the degree of temporal starvation of a link is as follows. Suppose that we choose a random point in time to observe a link $i$, and see how long it takes on average before link $i$ gets to transmit its next packet. We define this as the mean residual access time (MRAT) of the link. If the MRAT is excessive, we then said that link $i$ is starved. Let $Y$ be the random variable corresponding to the interval between two successive packet transmissions (i.e., inter-access time). Then, the probability density of the residual access time $X$ is $f_X(x) = (1 - F_Y(x))/E[Y]$, and the MRAT is given by [2]

$$\text{MRAT} = E[X] = \frac{E[Y^2]}{2E[Y]} \quad (25)$$

Fig. 8 plots MRAT versus $\rho$ for a link in the ring networks with $L = 1$ and $(q,k) = (1,1), (2,1)$ for $N = 16, 36, 64$. Fig. 9 plots the same for the case of $L = 2$ and $(q,k) = (1,1), (2,1),$ or $(3,1)$. In general, we see that frequency diversity can reduce the MRAT. However, even without frequency diversity, the MRAT of less than 10 packet transmission times in the (1,1) case is not excessive. Arguably, even real-time applications can be supported. Unfortunately, such is not the case with the 2D case.

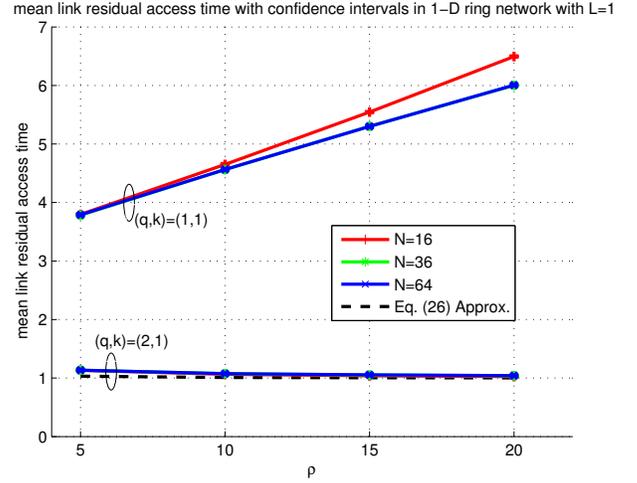

Fig. 8. MRAT versus access intensity for ring networks with $L = 1$, and $(q,k) = (1,1), (2,1)$.

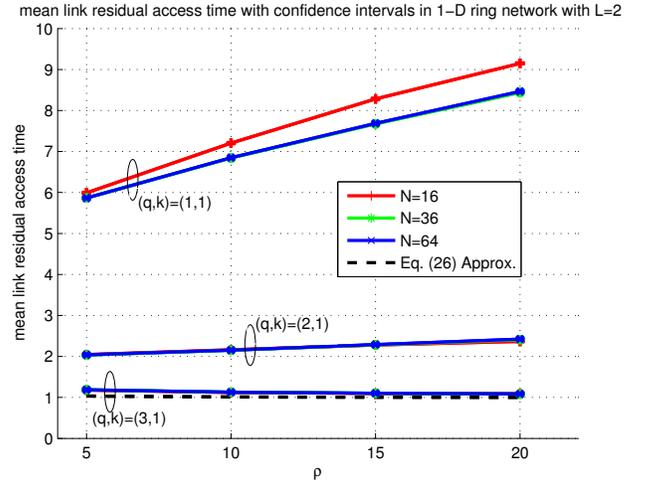

Fig. 9. MRAT versus access intensity for ring networks with $L = 2$, and $(q,k) = (1,1), (2,1), (3,1)$.

Fig. 10 plots MRAT versus $\rho$ for a link in the (1,1) torus. The MRAT increases exponentially with $\rho$ that we have to plot it on the log scale. As can be seen, the MRAT increases by two orders of magnitude to more than 1000 packet transmis-sion times as $\rho$ increases. Note that the equilibrium link throughput is near 0.5. Yet, once a link gets into a disadvantageous phase, it can be starved for a long time.

---
[2] We note that the first moment of inter-access time $E[Y]$ is not an appropriate metric for temporal starvation. The link throughput is basically $1/E[Y]$. As seen in the example of Fig. 7(b), temporal starvation may occur even when the stationary throughput is acceptable. Incorporating the second moment $E[Y^2]$ into the characterization of starvation was also considered in [19].

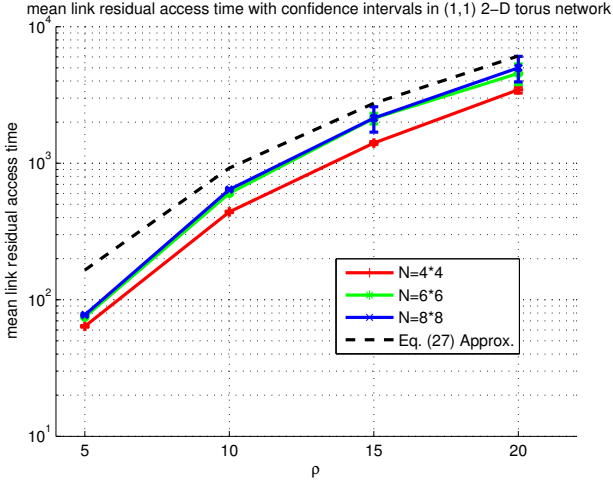

Fig 10. MRAT versus access intensity for torus networks with $(q,k)=(1,1)$.

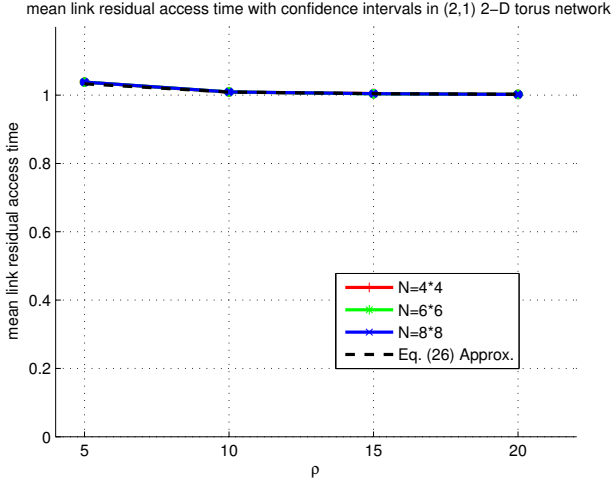

Fig. 11. MRAT versus access intensity for torus networks with $(q,k)=(2,1)$.

Fortunately, with the (2,1) torus, MRAT is reduced dramatically, as shown in Fig. 11. As mentioned earlier in Section IV-B, for large $\rho$, the situation faced by a link in the (2,1) torus is approximately the same as that faced by an isolated link. For an isolated link, the inter-access time $Y$ consists of the sum of two exponentially distributed random variables with mean $1/\rho$ and 1. Thus, $E[Y]=1/\rho+1$ and $E[Y^2]=2/\rho^2+2/\rho+2$. Substituting into (34), we get

$$\mathrm{MRAT}_{(2,1)\mathrm{torus}} \approx \mathrm{MRAT}_{\mathrm{isolated\ link}} = \frac{\rho^2+\rho+1}{\rho^2+\rho} \quad (26)$$

As indicated in Fig. 11, $\mathrm{MRAT}_{(2,1)\mathrm{torus}}$ is essentially the same as $\mathrm{MRAT}_{\mathrm{isolated\ link}}$ in (26) for all $\rho$ tested. Frequency diversity in the (2,1) torus makes links behave like isolated links.

Meanwhile, for the (1,1) torus, as mentioned in Section IV-B, for large $\rho$, half of the time, a link will behave like an isolated link, and the other half of the time, a link will behave like the center link of a five-link contention graph. The MRAT of the former is in (26). The MRAT of the latter can be obtained by transient analysis of a six-state birth-death process. Appendix D sketches the derivation and makes use of some standard results on passage times (e.g., Chapter 5 of [20]) to obtain the MRAT of the center link ((D2) of Appendix D). Overall, we have

$$\mathrm{MRAT}_{(1,1)\mathrm{torus}} \approx \frac{1}{2}\mathrm{MRAT}_{\mathrm{isolated\ link}} + \frac{1}{2}\mathrm{MRAT}_{\mathrm{center\ link\ of\ five-link\ star}}$$

$$= \frac{1}{2}\cdot\frac{\rho^2+\rho+1}{\rho^2+\rho} + \frac{1}{2}\cdot \quad (27)$$

$$\frac{12+108\rho+444\rho^2+924\rho^3+1156\rho^4+891\rho^5+429\rho^6+121\rho^7+15\rho^8}{12\rho(1+5\rho+6\rho^2+4\rho^3+\rho^4)}$$

As shown in Fig. 10, this "worst-case" analysis is an upper bound for the actual MRAT. As an approximation, it becomes more accurate as $\rho$ increases.

As a last observation, note from Fig. 8 and Fig. 9 that the isolated-link approximation (26) is also a very good approximation when $\rho$ is large for (2,1) ring with $L=1$ (2-colorable graphs for the $N$ values studied), and for (3,1) ring with $L=2$ (3-colorable graph for the $N$ values studied). Together with the observation that (26) is a good approximation when $\rho$ large for (2,1) torus (2-colorable graph for the $N$ values studied), this suggests that when $q$ in a $(q,1)$ system is equal to the chromatic number of the contention graph, then the links will decouple from each other and behave roughly like isolated links. The next section proves a general coloring theorem to this effect for general network topologies.

## VI. A COLORING THEOREM FOR FREQUENCY-AGILE CSMA NETWORKS

In this section, we consider $(q,1)$ networks with general topology but finite number of nodes. We argue that if the corresponding contention graph is $q$-colorable, then all vertices can have throughput approaching 1 when $\rho$ is large, and the probability of a link not transmitting at any given moment in time approaches zero.

**Definition 1:** A coloring of a graph is an assignment of colors to the vertices such each vertex and its neighbors have different colors. An uncolored vertex does not impose any constraint on its neighbors. A $q$-colorable graph is a graph in which all vertices can be successfully colored with a total of $q$ colors.

**Theorem 1:** Consider a $(q,1)$ frequency-agile CSMA network with finite number of nodes. If the contention graph is $q$-colorable, then $\lim_{\rho\to\infty}Th_i(\rho)=1$ and $\lim_{\rho\to\infty}\Pr[s_i=0]=0$ (i.e., the probability of a link $i$ not transmitting at any given moment in time approaches 0)

**Proof:** In a $(q,1)$ network, the probability of a state $s=(s_1,...,s_N)$ is $P_s=\rho^{n_s}/Z$, where $n_s$ is the number of links transmitting while the system is in state $s$, and $N$ is the total number of links (see (1) and (2)). Each state has an implied channel assignment to the transmitting links. Thus, we can specify each feasible state $s$ as a channel-assignment state CS: $\{(i, f_i)\}$, where $i$ is a vertex that is transmitting under state $s$, and $f_i$ is the channel on which it transmits.

Different CSs may have different numbers of transmitting links. We will call those with the maximum number of transmitting links maximum channel states (MCS). From the expression $P_s = \rho^{n_s}/Z$, it is obvious that when $\rho$ is large, the probabilities of the MCSs will dominate. To the extend that $\rho \to \infty$, only the MCSs will have appreciable probabilities, and all MCSs have equal probability.

Now, if the contention graph is $q$-colorable and we have a $(q,1)$ network, the cardinalities of all MCSs will be $N$, with each MCS corresponding to one way to color all vertices. In other words, all links will be transmitting while the system is in an MCS. Thus, $\lim_{\rho \to \infty} Th_i(\rho) = 1$ and $\lim_{\rho \to \infty} \Pr[s_i = 0] = 0$ □

*Comments on Theorem 1*

Although the above proof is simple, it relies on a non-trivial result. Namely, it relies on the fact that underlying the frequency-agile MAC is a time-reversible dynamic that gives rise to the product-form equation (2) for the stationary probability, without which the short proof will not be possible.

*Implications and Ramifications of Theorem 1*

We now examine several implications and ramifications of Theorem 1:

- *Coloring without explicit coloring algorithm* − Coloring problem is NP-hard in general. That means we will encounter much difficulty if we use a centralized algorithm to assign frequency channels to links in a large network. With our frequency-agile MAC protocol (see Section II-B for details), each link determines the channel it will use in a dynamic, adaptive and distributed manner. Theorem 1 indicates that the MAC protocol manages to color the graph successfully in a distributed manner using only carrier-sensing information, not an explicit coloring algorithm.
- *High throughput and elimination of temporal starvation* − The throughputs of all links are 1. High throughput and maximal fairness are achieved at the same time. Also, $\Pr[s_i = 0] = 0$ means that a link is nearly transmitting all the time, and there is no temporal starvation.
- *Decoupling of links* − Essentially, what Theorem 1 says is that as $\rho \to \infty$, with sufficient colors, the links will be decoupled from each other and behave like isolated links. Even for modest $\rho$, as can be seen from our analytical and simulation results, the decoupling effect is present.
- *Use of multi-radio technology* − Given Theorem 1, suppose that multi-radio technology is available at each node. An issue is how to set $q$ to ensure colorability. That is, how many of the radios to turn on. The following argues that a distributed algorithm is available.

The contention graph of a CSMA network is subjected to geographical constraints. Specifically, only nodes within a carrier-sensing range cannot transmit on the same channel. The following proposition relates the contention graph with unit disk graph.

**Proposition 1:** Consider a CSMA network in 2D free space, in which the transmitters of all links use the same carrier-sensing range. The contention graph of the network can be modeled as a unit disk graph, with one unit being the carrier-sensing range.[3]

**Proof:** For each link, we represent it as a vertex at the exact position of its transmitter. As such, there is an edge between two vertices if and only if the distance between their transmitters is less than the carrier-sensing range. Consequently, the contention graph can be modeled as a unit disk graph. □

Consider a unit disk graph *G*. Let $W_G$ be the size of the maximum clique. From [24], the chromatic number $C_G$ (number of colors needed to color all vertices *G*) is upper bounded by $6W_G - 6$. Furthermore, $W_G$ can be solved by a polynomial algorithm [25] and a distributed algorithm is available [26].

- *What if not enough colors* − The proof of Theorem 1 also suggests what will happen when there are not enough colors to color the contention graph. In this case, when $\rho$ is large, the dominant states with appreciable probabilities are still the MCSs. Since the total system throughput under each MCS is higher than other CSs that are not MCSs, the $(q,1)$ network still achieves the best possible overall system throughput.

## VII. RELATED WORK AND FURTHER REMARKS

The ideal carrier-sensing network (ICN) model was investigated in [1] and [13]. In particular, [1] provided a thorough investigation and uncovered many interesting outstanding issues arising out of ICN, one of which is how to exploit frequency diversity to combat use starvation problems. In that respect, the current paper fills a gap.

Ref. [1] also proved that the stationary probability distribution in ICN is insensitive to the distributions of the packet transmission time and backoff countdown time given the ratio of their mean $\rho$ (e.g., in practical CSMA networks such as Wi-Fi, the packet duration and backoff time are not exponentially distributed). The current paper extends ICN to the multi-frequency case. Since the underpinnings of the equations are similar, the insensitivity property carries to the multi-frequency case.

The carrier-sensing relationship in ICN is modeled by a contention graph. The graph-based carrier-sensing model is essential for the time-reversibility property that leads to the important product-form solution in (1). Ref. [23] provides a realistic implementation for the graph-based carrier sensing model. The interference model assumed in [23] is the SINR model, in which the interference experienced by a node is the sum total of all the powers received from all simultaneous transmission. To ensure that the permitted simultaneous transmissions are interference-safe (i.e., the SINRs at all the receivers are above a target requirement), [23] proposed an incremental power carrier-sensing (IPCS) mechanism that monitors the power increments sensed in the recent past rather than just the absolute power sensed at the moment. This allows the graph-based carrier-sensing model in ICN to be realized, while ensuring interference-safe operation under the realistic, non-graph-based SINR model.

---

[3] Refs. [21], [22] and [23] showed that a CSMA network in 2D free space can be made to be hidden-node-free by setting a proper carrier-sensing range uniformly across all pairs of nodes.

We remark that absolute power carrier sensing will lead to a system without the time-reversibility property. One can easily come up with an example with three links, in which it is possible to go from simultaneous transmissions $\{1, 2\}$ to simultaneous transmissions $\{1, 2, 3\}$; but not from $\{1, 3\}$ to $\{1, 2, 3\}$, because link 2 senses a large absolute power from links 1 and 3. In this way, if the system moves from the state $\{1, 2, 3\}$ to the state $\{1, 3\}$, the reverse transition is not possible; hence the system is not time-reversible.

Ref. [12] investigated 1D and 2D regular networks based on the ICN model without providing closed-form solutions. As far as we know, our closed-form throughput solutions to 1D ring/linear CSMA networks making use of transfer matrices are new. Also new is the result that the 2D torus is equivalent to the 2D Ising model with external magnetic field, hence is most likely not amenable to exact analysis. Our simple approximate analysis for 2D torus with and without frequency diversity is also new.

There are two approaches to distributed dynamic frequency-channel assignment in CSMA networks: 1) the network-layer approaches [2-5]; and 2) the MAC-layer approaches [6-11]. For the network-layer approaches, after a node is assigned a frequency channel, its MAC layer will only execute the CSMA protocol on that channel and will be oblivious to what happens on other channels.

With recent advances in multi-radio technologies, it is becoming feasible and cost-effective for a node to operate on multiple frequency channels at a time. The MAC-layer approaches execute the CSMA protocol over multiple-frequency channel simultaneously. They can be a lot more adaptive and can respond more quickly to dynamic changes in network topology than the network-layer approaches. The frequency-agile MAC proposed in this paper is one possible design. Ref. [6] proposed a multi-channel MAC to mitigate "equilibrium" starvation. We believe the use of frequency diversity to address "temporal" starvation is new and has not been investigated before our paper.

An advantage of our design, besides its good throughput and robustness against temporal starvation, is that its dynamic can be characterized by equations and be subjected to rigorous analysis. An example is our coloring theorem in Section VI, which arises naturally way out of the ICN equation. Another example is that the analytical framework of our frequency-agile MAC also provides a foundation for further work, including investigations of frequency-agile networks in which different links may adopt different access intensities in a dynamic manner [2][14].

## VIII. CONCLUSION

The paper proposes and analyzes the performance of a simple frequency-agile MAC protocol for CSMA networks. We focus on throughput and temporal starvation. The latter is an important concern because the traditional CSMA protocol is susceptible to temporal starvation, in which even links that have acceptable equilibrium throughput can still can receive near-zero throughput for very long durations [1][12][27]. While there have been many investigations on boosting equilibrium throughput in CSMA networks, to our best knowledge, this paper is a first attempt to explore solving the temporal starvation problem using frequency diversity.

An advantage of our frequency-agile MAC is that the dynamic of a link using the protocol can be modeled precisely so that the performance of the overall network can be analyzed rigorously. For our analysis, we focus on 1D and 2D regular networks, and provide closed-form solutions and approximations for link throughput and mean residual access time.

We augment the analysis of the regular networks with a coloring theorem applicable to networks of general topology. Together, they imply that with enough frequency channels, our frequency-agile MAC can decouple the detrimental interactions between neighboring links to increase throughput and robustness against temporal starvation. Notably, our MAC protocol achieves this in a distributive and adaptive manner without executing an "explicit" coloring algorithm for channel assignment.

## IX. ACKNOWLEDGEMENTS

We thank C. H. Kai for sharing her simulation codes for ICN, which we extend for our simulation study here; drafting of some figures in this paper; and working out the solution to a set of linear equations that yield (D2).

## References

[1] S. C. Liew, C. H. Kai, H. C. Leung, P. Wong, "Back-of-the-Envelope Computation of Throughput Distributions in CSMA Wireless Networks", *IEEE ICC*, June 09; long version *to appear in IEEE Trans Mobile Computing*, Sep 2010. Technical report available at http://arxiv.org//pdf/0712.1854.

[2] M. Chen, S. C. Liew, Z. Shao, C. Kai, "Markov Approximation for Combinatorial Network Optimization," *IEEE Infocom*, 2010.

[3] A. Mishra, S. Banerjee, W Arbaugh, "Weighted Coloring based Channel Assignment for WLANs," *ACM SIGMOBILE Mobile Comp. and Commun. Rev.*, vol. 9, no. 3, 2005.

[4] B-J Ko, V. Misra, J. Padhye, D. Rubenstein, "Distributed Channel Assignment in Multi-radio 802.11 Mesh Networks," *IEEE WNCN 2007*.

[5] A. Mishra, V. Shrivastava, D. Agrawal, S. Banarjee, S. Ganguly, "Distributed Channel Management in Uncoordinated Wireless Environment," *ACM Mobicom 06*.

[6] J. Shi, T. Salonidis, E. W. Knightly, "Starvation Mitigation through Multi-channel Coordination in CSMA Multi-hop Wireless Networks," *ACM Mobicom 06*.

[7] A. Nasipuri, J. Zhang, S. R. Das, "A Multichannel CSMA MAC Protocol for Multihop Wireless Networks," *IEEE WNCN 99*.

[8] S-L Wu, C-Y Lin, Y-C Tseng, J-P Sheu, "A New Multi-Channel MAC Protocol with On-Demand Channel Assignment for Multi-hop Mobile Ad Hoc Networks*,"* IEEE I-SPAN 2000.
[9] R. Maheshwari, H. Gupta, S. R. Das, "Multichannel MAC Protocols for Wireless Networks," *IEEE SECON 2006.*
[10] J. So, N. Vaidya, "Multi-Channel MAC for Ad Hoc Networks: Handling Multi-Channel Hidden Terminals using a Single Transceiver," *ACM Int. Symp.on Mobile and Ad Hoc Networking and Computing, 04.*
[11] J. Mo, H-S W. So, J. Walrand, "Comparison of Multichannel MAC Protocols," *IEEE Trans. Mobile Computing*, vol. 7, no. 1, 2008.
[12] M. Durvy, O. Dousse, P. Thiran, "Border Effects, Fairness, and Phase Transition in Large Wireless Networks," *IEEE Infocom 08*.
[13] X. Wang and K. Kar, "Throughput Modeling and Fairness Issues in CSMA/CA Based Ad hoc networks," *IEEE Infocom, 05*.
[14] L. Jiang, J. Walrand, "A Distributed CSMA Algorithm for Throughput and Utility Maximization in Wireless Networks," to appear in *IEEE/ACM Trans on Networking*, 2010.
[15] IEEE Std 802.11-1997, IEEE 802.11 Wireless LAN Medium Access Control (MAC) and Physical Layer (PHY) Specifications.
[16] Independent set, http://en.wikipedia.org/wiki/Independent_set.
[17] S. Kakumanu, R. Sivakumar, "Gila: A Practical Solution for Effective High Data Rate WiFi-Array," *ACM Mobicom 09.*
[18] Ising model, http://en.wikipedia.org/wiki/Ising_model.
[19] S. C. Liew, Y. J. Zhang and D. R. Chen, "Bounded Mean-Delay Throughput and Non-Starvation Conditions in Aloha Network," *IEEE Trans. Networking*, vol.17, no. 5, 2009.
[20] J. Keilson, *Markov Chain Models-Rarity and Exponentiality*, Springer-Verlag, 1979.
[21] L. B. Jiang, S. C. Liew, "Hidden-node Removal and Its Application in Cellular WiFi Networks," *IEEE Trans.Vehicular Technology*, vol. 56, no. 5, 2007.
[22] C. K. Chau, M. Chen, S. C. Liew, "Capacity of Large Scale CSMA Wireless Networks", *ACM Mobicom 09*.
[23] L. Fu, S. C. Liew, J. Huang, "Effective Carrier Sensing in CSMA Networks under Cumulative Interference," *IEEE Infocom*, 2010.
[24] A. Graf, M. Stumpf, G. Weienfels, "On coloring unit disk graphs", *Algorithmica*, vol. 20, no. 3, 1998.
[25] B. N. Clark, C. J. Colbourn, D. S. Johnson , "Unit Disk Graphs", *Discrete Mathematics*, vol 86, 1990.
[26] E. Jennings and L.Motycková, "A distributed algorithm for finding all maximal cliques in a network graph" *1st Latin American Symposium on Theoretical Informatics 92*.
[27] L. Fu, S. C. Liew, J. Huang, "Effective Carrier Sensing in CSMA Networks under Cumulative Interference," *IEEE Infocom*, 2010.
[28] C. H. Kai and S. C. Liew, "Temporal Starvation in CSMA Wireless Networks," Technical Report, The Chinese University of Hong Kong, 2010.

## APPENDIX A: ANALYSIS OF 1-D LINEAR NETWORK

We consider the case of $(q,k) = (1,1)$. The same analytical approach is applicable to other cases. The open-ended linear network with $N$ vertices is obtained by breaking the edge between vertices 1 and $N$ in the ring network. The partition function is still of the form $Z = \sum_{s_1} s_1^T \mathbf{A}^N s_{N+1}$ except that $s_{N+1} \neq s_1$, and $s_{N+1}$ does not impose any restriction on the network. Specifically, $s_{N+1} = (1 \ 0)^T$ (i.e., this is equivalent to an $(N+1)$-vertex network in which vertex $N+1$ is always in the idle state). Thus,

$$Z = \sum_{s_1} s_1^T \mathbf{A}^N \begin{pmatrix} 1 \\ 0 \end{pmatrix} = (1 \ 1) \mathbf{A}^N \begin{pmatrix} 1 \\ 0 \end{pmatrix} \quad (A1)$$

where $\mathbf{A}$ is the same as that in the ring network as in (5). Let us define $a_{gh}^{(i)}$ such that

$$\mathbf{A}^i \triangleq \begin{pmatrix} a_{00}^{(i)} & a_{01}^{(i)} \\ a_{10}^{(i)} & a_{11}^{(i)} \end{pmatrix}, \quad i \in \{1,...,N\} \quad (A2)$$

By diagonalization of $\mathbf{A}^i$, it can be shown that

$$\mathbf{A}^i = \frac{1}{z_1 - z_2} \begin{pmatrix} z_1^{N+1} - z_2^{N+1} & z_1^N - z_2^N \\ -z_2 z_1^{N+1} + z_1 z_2^{N+1} & -z_2 z_1^N + z_1 z_2^N \end{pmatrix} \quad (A3)$$

Substitution into (A1) gives

$$Z = a_{00}^{(N)} + a_{10}^{(N)} = \frac{z_1^{N+2} - z_2^{N+2}}{z_1 - z_2} \quad (A4)$$

where $z_1$ and $z_2$ are eigenvalues of $\mathbf{A}$ given in (7) (note that $1 - z_2 = z_1$ and $1 - z_1 = z_2$).

In the open-ended linear network, different vertices have different throughputs and we cannot use $Th_i = (\rho d \ln Z/d\rho)/N$ to obtain the throughput of vertex $i$.

Consider vertex $i$. The marginal probability $P_{s_i}$ can be obtained by fixing the value of $s_i$ in (4) while summing over all values $s_1,...,s_{i-1}, s_{i+1},...,s_N$. Doing so gives

$$P_{s_i} = \frac{(1 \ 1) A^{i-1} s_i s_i^T A^{N+1-i} \begin{pmatrix} 1 \\ 0 \end{pmatrix}}{Z} = \frac{\left(a_{0 s_i}^{(i-1)} + a_{1 s_i}^{(i-1)}\right) a_{s_i 0}^{(N+1-i)}}{a_{00}^{(N)} + a_{10}^{(N)}} \quad (A5)$$

In particular, the throughput of vertex $i$ is the "on" probability given by setting $s_i = 1$ in the above:

$$Th_i(\rho, N) = \frac{-z_1 z_2 (z_1^i - z_2^i)(z_1^{N+1-i} - z_2^{N+1-i})}{(z_1 - z_2)(z_1^{N+2} - z_2^{N+2})} \quad \forall \ i \in \{1,...,N\} \quad (A6)$$

For large $N$, the larger eigenvalue $z_1$ dominates. We have

$$Th_i(\rho, \infty) \approx \frac{-z_2 \left(z_1^i - z_2^i\right)}{(z_1 - z_2) z_1^i} \quad \forall \ i \in \{1,...,\lceil N/2 \rceil\} \quad (A7)$$

At the edge and the middle of the graph, we have

$$Th_1(\rho, \infty) = \frac{-z_2}{z_1} = \frac{\sqrt{1+4\rho}-1}{\sqrt{1+4\rho}+1}$$

$$Th_{\lceil N/2 \rceil}(\rho, \infty) = \frac{-z_2}{(z_1 - z_2)} = \frac{\sqrt{1+4\rho}-1}{2\sqrt{1+4\rho}} \quad (A8)$$

Indeed, $Th_i(\rho, \infty) \approx \left(\sqrt{1+4\rho}-1\right)/\left(2\sqrt{1+4\rho}\right)$ for sufficiently large $i$ that $z_1^i$ dominates over $z_2^i$. That is, as we move from the edge of the graph to the inside, all vertices have roughly the same throughput; furthermore, this internal throughput is the same as the throughput of a vertex in symmetric ring network, as in (10). Note also that the throughput at the edge is larger than inside. For large $\rho$, the throughput at the edge approaches 1 while that inside approaches $1/2$.

## APPENDIX B: THROUGHPUT ANALYSIS OF THIN-STRIP 2D NETWORKS

This appendix presents the transfer-matrix analysis for the thin-strip network shown in Fig. 3(b).

(1,1) *Case:* For a local unit $i$, $s_i = (1\ 0\ 0)^T$ if none of its two vertices transmits; $s_i = (0\ 1\ 0)^T$ if the top vertex transmits; $s_i = (0\ 0\ 1)^T$ if the bottom vertex transmits. The transfer matrix is

$$\mathbf{A} = \begin{pmatrix} 1 & 1 & 1 \\ \rho & 0 & \rho \\ \rho & \rho & 0 \end{pmatrix} \quad (B1)$$

We note that the state structure $s_i$ and $\mathbf{A}$ are exactly the same of that of the (2,1) 1-D ring network, as in (11). Therefore, the analyses of both cases are the same except for the interpretation of the result. Over there, each unit consists of one vertex that operates on two channels; here, each unit consists of two vertices and there is only one channel. We only need to divide the vertex throughputs in (15) by two to obtain the vertex throughput here.

(2,1) *ICN Case:* Now, consider the (2,1) case for the network in Fig. 3(b). Each unit has seven possible states. Let the state of unit $i$ be $s_i \in \{00, 01, 10, 02, 20, 12, 21\}$, where the first number in $s_i$ is the state of the top vertex and the second number is the state of the bottom vertex in the unit. For example, $s_i = 01$ means the top vertex is idle and the second vertex is transmitting on channel 1; $s_i = 12$ means the top vertex is transmitting on channel 1 and the bottom vertex is transmitting on channel 2. The transfer matrix is

$$\mathbf{A} = \begin{array}{c} \\ 00 \\ 01 \\ 10 \\ 02 \\ 20 \\ 12 \\ 21 \end{array} \begin{pmatrix} 00 & 01 & 10 & 02 & 20 & 12 & 21 \\ 1 & 1 & 1 & 1 & 1 & 1 & 1 \\ \rho & 0 & \rho & \rho & \rho & \rho & 0 \\ \rho & \rho & 0 & \rho & \rho & 0 & \rho \\ \rho & \rho & \rho & 0 & \rho & 0 & \rho \\ \rho & \rho & \rho & \rho & 0 & \rho & 0 \\ \rho^2 & \rho^2 & 0 & 0 & \rho^2 & 0 & \rho^2 \\ \rho^2 & 0 & \rho^2 & \rho^2 & 0 & \rho^2 & 0 \end{pmatrix} \quad (B2)$$

The characteristic polynomial is of the $7^{th}$ order. The dominant eigenvalue is approximately $z_1 = \rho^2 + 2\rho + 4$ for large $\rho$. The vertex throughput for large $\rho$ and $N$ is

$$Th_i(\rho, \infty) = \frac{\rho(d\ln Z/d\rho)}{2N} \approx \frac{\rho dz_1/d\rho}{2z_1} = 1 - \frac{\rho + 2}{\rho^2 + 2\rho + 2} \quad (B3)$$

## APPENDIX C: CORRESPONDENCE OF ISING MODELS WITH EXTERNAL MAGNETIC FIELD AND CSMA NETWORKS

*1-D Ring*

For 1-D ring Ising model with external magnetic field, we have [18]

$$P_{(s_1',s_2',...,s_N')} = \frac{\exp\left[\sum_{i=1}^{N}(hs_i' + Ks_i's_{i+1}')\right]}{\sum_{\substack{s_1',s_2',...,s_N' \\ s_i' \in \{-1,+1\}}} \exp\left[\sum_{i=1}^{N}(hs_i' + Ks_i's_{i+1}')\right]} \quad (C1)$$

where $h$ is the external magnetic field, and $K$ is the ferromagnetic interaction. We want to map the state of the Ising model, $(s_1', s_2', ..., s_N')$, $s_i' \in \{-1, +1\}$, to the state of a CSMA ring network, $(s_1, s_2, ..., s_N)$, $s_i \in \{0, 1\}$. Define $s_i = (s_i' + 1)/2$. Substituting into (C1), we get

$$P_{(s_1,s_2,...,s_N)} = \frac{\exp\left\{\sum_{i=1}^{N}[(2h-4K)s_i + 4Ks_is_{i+1}]\right\}}{\sum_{\substack{s_1,s_2,...,s_N \\ s_i \in \{0,1\}}} \exp\left\{\sum_{i=1}^{N}[(2h-4K)s_i + 4Ks_is_{i+1}]\right\}} \quad (C2)$$

In the limit that $h, K \to -\infty$, but $2h - 4K = r$, the above becomes

$$P_{(s_1,s_2,...,s_N)} = \frac{\prod_{i=1}^{N} \psi(s_i, s_{i+1})\exp(rs_i)}{\sum_{\substack{s_1,s_2,...,s_N \\ s_i \in \{0,1\}}} \prod_{i=1}^{N} \psi(s_i, s_{i+1})\exp(rs_i)} \quad (C3)$$

where $\psi(s_i, s_{i+1}) = 0$ if $s_i = s_{i+1} = 1$, and $\psi(s_i, s_{i+1}) = 1$ otherwise. Eqn. (C3) is simply the equation for the 1-D ring CSMA network with $\phi_i(s_i) = \rho^{s_i} = \exp(rs_i)$ in (4)

*2-D Torus*

The correspondence for the 2-D torus case is similar. The state variables $s_i$ and $s_i'$ are replaced by the state variables $s_{i,j}$ and $s_{i,j}'$, respectively. The Ising model equation (C1) becomes

$$P_{s'} = \frac{\exp\left[\sum_{i=1}^{N}\sum_{j=1}^{N}(hs_{i,j}' + Ks_{i,j}'s_{i+1,j}' + Ks_{i,j}'s_{i,j+1}')\right]}{\sum_{\forall s' \in \mathcal{X}'^{N^2}} \exp\left[\sum_{i=1}^{N}(hs_{i,j}' + Ks_{i,j}'s_{i+1,j}' + Ks_{i,j}'s_{i,j+1}')\right]} \quad (C4)$$

where $\mathcal{X}' \in \{-1, +1\}$. With transformation $s_{i,j} = (s_{i,j}' + 1)/2$, (C4) becomes

$$P_s = \frac{\exp\left\{\sum_{i=1}^{N}\sum_{j=1}^{N}\left[(2h-8K)s_{i,j} + 4Ks_{i,j}s_{i+1,j} + 4Ks_{i,j}s_{i,j+1}\right]\right\}}{\sum_{\forall s \in \mathcal{X}^{N^2}} \exp\left\{\sum_{i=1}^{N}\sum_{j=1}^{N}\left[(2h-8K)s_{i,j} + 4Ks_{i,j}s_{i+1,j} + 4Ks_{i,j}s_{i,j+1}\right]\right\}} \quad (C5)$$

where $\mathcal{X} \in \{0, 1\}$. In the limit that $h, K \to -\infty$, but $2h - 8K = r \triangleq \ln\rho$, the above becomes

$$P_s = \frac{\prod_{i=1}^{N}\prod_{j=1}^{N}\psi(s_{i,j},s_{i+1,j})\psi(s_{i,j},s_{i,j+1})\exp(rs_{i,j})}{\sum_{\forall s\in\mathcal{X}^{N^2}}\prod_{i=1}^{N}\prod_{j=1}^{N}\psi(s_{i,j},s_{i+1,j})\psi(s_{i,j},s_{i,j+1})\exp(rs_{i,j})} \quad (C6)$$

which is the equation for the 2-D torus CSMA network.

## APPENDIX D: DERIVATION OF THE MEAN RESIDUAL ACCESS TIME OF THE CENTER VERTEX OF A FIVE-VERTEX CONTENTION GRAPH.

Consider a star contention graph consisting of five vertices, with one vertex in the center being flanked by four vertices. We are interested in the mean residual access time of the vertex in the middle. Let $Y$ denote the inter-access time (i.e., the interval between the beginnings of two successive transmissions) of the center vertex. It is basically the recurrent time of the active state of the center vertex. If we can obtain $E[Y]$ and $E[Y^2]$, we can then obtain $\text{MRAT} = E[Y^2]/(2E[Y])$.

We can model the experience of the center vertex by a six-state birth-death process. The six states consist of a state $c$, during which the center vertex transmits; an idle state $i$ during which nobody transmits; and states 1, 2, 3, and 4, during which one, two, three, and four of the outer links transmit, respectively. Fig. D1 shows the birth-death process, in which one time unit is normalized to be the mean packet transmission time.

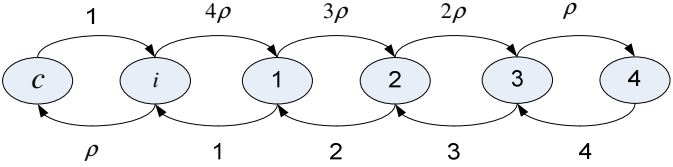

Fig. D1. Six-state birth-death process modeling the experience of the center vertex of a five-vertex star contention graph.

We have $Y = T_{c\to i} + T_{i\to c}$, where $T_{c\to i}$ is the time for the transition $c\to i$ to occur given that the system is in state $c$, and $T_{i\to c}$ is the passage time from state $i$ to state $c$. Note that $T_{c\to i}$ is exponentially distributed with $E[T_{c\to i}]=1$ and $E[T^2_{c\to i}]=2$.

Now,

$$\begin{aligned}\text{MRAT} &= \frac{E[Y^2]}{2E[Y]} = \frac{E[(T_{c\to i}+T_{i\to c})^2]}{2E[T_{c\to i}+T_{i\to c}]}\\ &= \frac{E[T^2_{c\to i}]+E[T^2_{i\to c}]+2E[T_{c\to i}]E[T_{i\to c}]}{2(E[T_{c\to i}]+E[T_{i\to c}])}\\ &= \frac{2+Var[T_{i\to c}]+E^2[T_{i\to c}]+2E[T_{i\to c}]}{2(1+E[T_{i\to c}])}\end{aligned} \quad (D1)$$

Thus, the whole derivation of the MRAT boils down to finding $E[T_{i\to c}]$ and $Var[T_{i\to c}]$ in the birth-death process. We use the formulas (5.2.4) and (5.2.5) in [20] to crunch out $E[T_{i\to c}]$ and $Var[T_{i\to c}]$. Plugging them into (D1) yields

$$\text{MRAT} = \frac{12+108\rho+444\rho^2+924\rho^3+1156\rho^4+891\rho^5+429\rho^6+121\rho^7+15\rho^8}{12\rho(1+5\rho+6\rho^2+4\rho^3+\rho^4)} \quad (D2)$$